\documentclass[pre,superscriptaddress,showkeys,showpacs,twocolumn]{revtex4-1}
\usepackage{graphicx}
\usepackage{latexsym}
\usepackage{amsmath}
\begin {document}
\title {Generic criticality of community structure in random graphs}
\author{Adam Lipowski}
\affiliation{Faculty of Physics, Adam Mickiewicz University, Pozna\'{n}, Poland}
\author{Dorota Lipowska}
\affiliation{Faculty of Modern Languages and Literature, Adam Mickiewicz University, Pozna\'{n}, Poland}
\begin {abstract}
We examine a community structure in random graphs of size $n$ and link probability $p/n$ determined with the Newman greedy optimization of modularity. Calculations show that  for $p<1$ communities are nearly identical with clusters. For $p=1$ the average sizes of a community $s_{av}$ and of the giant community $s_g$ show a power-law increase  $s_{av}\sim n^{\alpha'}$ and $s_g\sim n^{\alpha}$. From numerical results we estimate $\alpha'\approx 0.26(1)$, $\alpha\approx 0.50(1)$, and using the probability distribution of sizes of communities we suggest that $\alpha'=\alpha/2$ should hold. For $p>1$ the community structure remains critical: (i) $s_{av}$ and $s_g$ have a power law increase with $\alpha'\approx\alpha <1$; (ii) the probability distribution of sizes of communities is very broad and nearly flat for all sizes up to $s_g$. For large $p$ the modularity $Q$ decays as $Q\sim p^{-0.55}$, which is intermediate  between some previous estimations.    
To check the validity of the results, we also determined the community structure  using another method, namely a non-greedy optimization of modularity. Tests with some benchmark networks show that the method outperforms the greedy version. For random graphs, however, the characteristics of the community structure determined using both greedy an non-greedy optimizations are, within small statistical fluctuations, the same.
 \end{abstract}

\maketitle
\section{Introduction}
Random graphs are unique structures, which are of interest both for mathematical and physical sciences. They allow a deep rigorous analysis \cite{erdos} and moreover provide an excellent testing ground for various methods developed in the context of complex networks \cite{barabasi,mendes}. One of the most basic features of random graphs is their percolative properties related to the formation of clusters of various sizes. It is well known that a large graph of $n$ nodes with the link probability $p/n$  has an interesting transition at $p=1$. For $p<1$ the largest cluster (giant component) is relatively small and its size $s$ slowely increases with the size of the graph ($s\sim \log(n)$). However, for $p>1$ the giant component spans a finite fraction of the graph ($s\sim n$). At the transition point the size of the giant component is intermediate between these two regimes and $s\sim n^{2/3}$ \cite{bollobas}. Instead of clusters of links one might consider clusters of more complex objects, e.g., the so-called $k$-cliques that are complete subgraphs of $k$ vertices. For $k$-cliques, interesting percolation transitions were also identified \cite{vicsek}.

Subgraphs having strong internal connections, of which $k$-cliques are extreme cases, play an important role in the functioning of many real networks. The existence of such structures, often called communities, indicates a hierarchical organization of networks and influences their stability and robustness \cite{fortunato}. However, finding the community structure, despite considerable interest, is not a well-defined problem---mainly for lack of a commonly accepted definition of the community. For example, restricting the notion of a community to a k-clique  is too stringent in many cases and for networks where such structures are sparse might lead to meaningless results.  Some other definitions of a community were proposed based, for example, on the ratio of inter- and intra-community links, but they contain some degree of  arbitrariness. In some cases no precise definition of a community is used and the partition into communities is simply an outcome of a given algorithm.
As a result, a multitude of methods are used to determine the community structure \cite{fortunato,boccaletti}.

\begin{figure}
\includegraphics[width=\columnwidth]{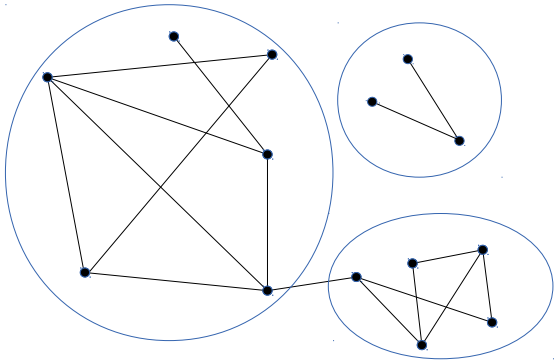}  \vspace{-0.8cm} 
\caption{Two clusters (separate subgraphs) and three communities (in circles): clusters might include one or more communities -- and different clusters incorporate different communities. 
\label{communities}}
\vspace{-0.0cm}
\end{figure}

It is natural to expect some kind of relation between clusters and communities. While for clusters the connectedness is a sufficient condition, for communities it is only a necessary one (Fig.~\ref{communities}). Communities are thus not larger than clusters, but a more precise comparison between these two basic structures seems to be missing. Such comparison would be feasible especially for random graphs, since their cluster properties, as we described above, are to a large extent known exactly. Particularly interesting might be to check whether the transition at $p=1$, which leads to the formation of the percolating cluster, induces a similar change in the community structure. Such an analysis is the main objective of the present paper.  

At first sight, the idea to examine a community structure in random graphs might appear questionable  on very basic grounds. Namely, random graphs almost by definition should not have any community structure. However, the problem is more subtle. Random graphs remain homogeneous (and thus structureless) but only on average. Any specific configuration of random graphs is subject to fluctuations and some communities might and do form \cite{Lancichinetti,amaral}. 

Most of our results are obtained using the Newman greedy optimization of modularity, which we briefly describe in Section II. In Section III we describe our numerical results obtained using this method and in particular we show that for $p\geq 1$ the community structure remains critical. To examine the validity of our approach, we also determined the community structure using a non-greedy optimization of modularity (section IV). Testing this method with some benchmark networks (Karate club, 4-module), we found that it outperforms the greedy optimization. The method preserves the basic algorithmic structure of its greedy counterpart and we thus expected that it might be used to examine a community structure in large networks. The non-greedy optimization applied to random graphs returns, however, nearly the same results as the greedy version. Such an agreement suggests that the criticality of community strucure is an inherent property of random graphs. Section V contains our conclusions.

\section{Greedy Optimization of Modularity}
To identify communities in random graphs, we used Newman's method \cite{newman2004} based on the optimization of the modularity $Q$ \cite{newmanq,newmanbook} defined as:
\begin{equation}
Q = \sum_i (e_{ii}-a_i^2),
\label{eq:modularity}
\end{equation}
where $e_{ii}$ is the fraction of links that have both ends in the community $i$, $a_i$ is the fraction of ends of links that are attached
to sites in community $i$, and the summation in Eq.~(\ref{eq:modularity}) is over communities.
The algorithm starts with all single-site communities and then successively amalgamates them in larger ones, choosing at each step the pair of communities the amalgamation of which gives the biggest possible increase in modularity (or the smallest decrease if no choice gives an increase). Eventually, all sites form a single community, but typically on the way there is a required solution: a configuration of communities with the largest modularity.
Similarly to other greedy algorithms, the obtained solution is usually only approximately optimal but the advantage of this  method is its computational efficiency, due to which it can be applied to very large networks. 

Implementing some of the modifications proposed by Clauset {\it et al.}  \cite{clauset} and adapting this frequently used method for sparse graphs, we were able to efficiently determine the community structure of random graphs with $n$ up to $3\cdot 10^6$ sites. To compare their cluster and community structure properties, we calculated the average size of the giant community  $s_g$ and the average size of the community $s_{av}$.
Provided that the algorithm results in the decomposition of the graph into communities of sizes $s_i$, where $i=1, 2,\ldots, l$ and $l$ is the number of the communities found in a given graph, the average size of the community is calculated as $s_{av}=\langle \frac{1}{n}\sum_{i=1}^l s_i^2\rangle$, where $<\ldots>$ stands for the average over independently generated graphs. 
\section{numerical results}
In the present section we describe our results obtained by the greedy optimization method. However, as we will show in the next section, for random graphs this method most likely returns nearly optimal solutions. 

\begin{figure}
\includegraphics[width=\columnwidth]{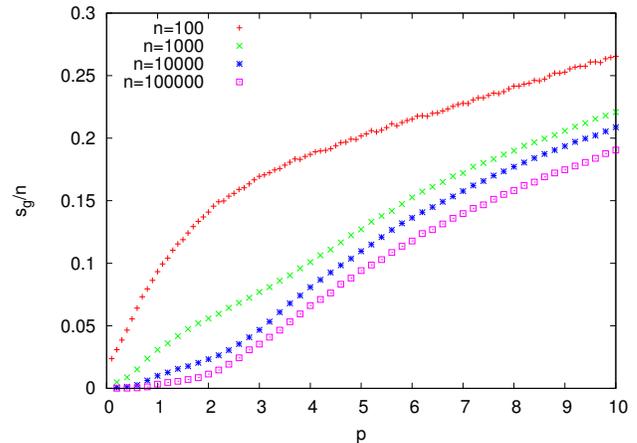}  \vspace{-0.8cm} 
\caption{(Color online) The rescaled size of the giant community $s_g/n$ as a function of the link probability $p$ calculated for several sizes of a graph $n$. For each $p$ and $n$ we average over $10^3-10^5$ graphs.
\label{giant-pn}}
\vspace{-0.0cm}
\end{figure}

In Fig.~\ref{giant-pn} we present the rescaled size of the giant component $s_g/n$ as a function of $p$. For $p<1$, $s_g/n$ quickly converges to zero, which indicates that the giant community, similarly to the giant component, contains a negligible fraction of sites. For $p>1$ the giant community seems to be much larger but a substantial size-dependence is also clearly visible.

\begin{figure}
\includegraphics[width=\columnwidth]{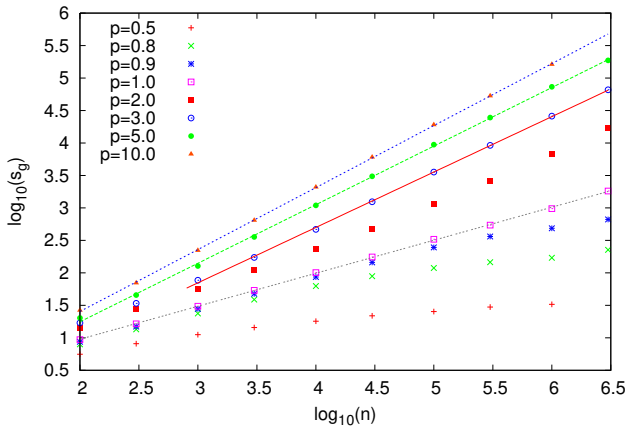}  \vspace{-0.8cm} 
\caption{(Color online)  The $n$-dependence of the size of the giant community  (log-log scale). For \mbox{$p\geq1$}, the size of the giant community shows a power-law increase $s_g\sim n^{\alpha}$ and  from the fit to numerical data we obtain $\alpha$(p=1)=0.51(1), $\alpha$(p=3)=0.85(1), $\alpha$(p=5)=0.90(1), and $\alpha$(p=10)=0.95(1). Let us notice that the numerical data show a nearly linear increase for more than 4 decades of $n$.
\label{log-giant}}
\vspace{-0.0cm}
\end{figure}

Let us notice that for $p>1$ the giant component contains a finite fraction of sites \cite{erdos}. Fig.~\ref{log-giant} shows that the giant community might not be that large. Indeed, even up to $p=10$ the asymptotic increase of $s_g$ obeys the power law $s_g\sim n^{\alpha}$ but $\alpha$ remains smaller than 1. It means that the giant community increases slower than the size of the graph $n$ and in the limit $n\rightarrow\infty$ it spans only a vanishingly small fraction of the graph. For $p<1$ our results show a slower than power-law increase of $s_g$ and it might be similar to the behaviour of the giant component that in the subcritical phase increases only logrithmically with $n$. At $p=1$, which is the interface of these two regimes, our results show that $s_g\sim n^{0.51}$. It is possible that in this case $s_g\sim n^{1/2}$ and it would be desirable to provide analytical arguments for such increase.  Let us recall that at $p=1$ the giant component is known to scale as $n^{2/3}$ \cite{bollobas}.

\begin{figure}
\includegraphics[width=\columnwidth]{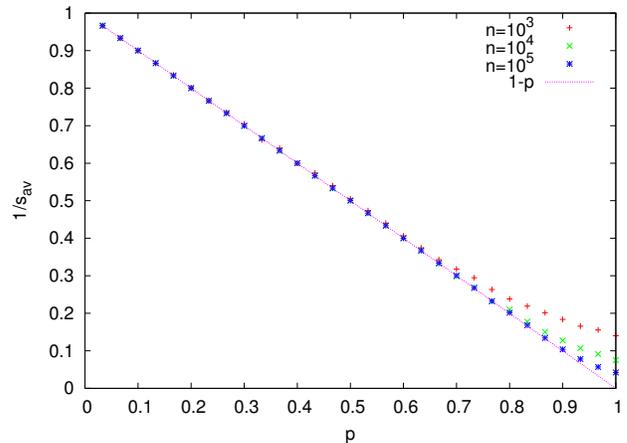}  \vspace{-0.8cm} 
\caption{(Color online) The inverse of the average community size $s_{av}$ as a function of $p$. For random graphs the average cluster size equals $1/(1-p)$ \cite{newmanavsize} and our numerical data  in the limit $n\rightarrow{\infty}$ are in a very good agreement with this formula. It means that for $p<1$ communities basically coincide with clusters.
\label{avsize}}
\vspace{-0.0cm}
\end{figure}

\begin{figure}
\includegraphics[width=\columnwidth]{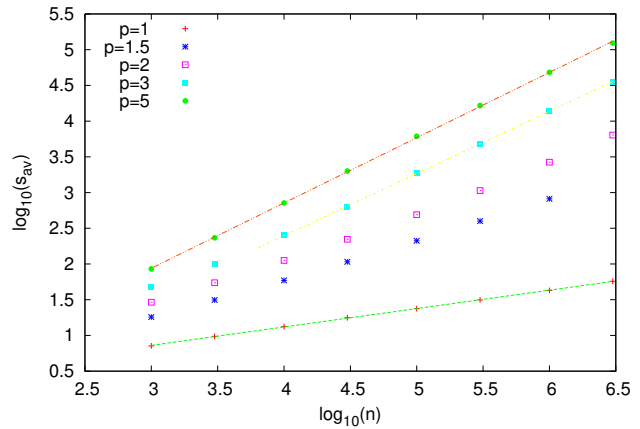}  \vspace{-0.8cm} 
\caption{(Color online)   The $n$-dependence of the average size of a community  $s_{av}$ (log-log scale). For $p\geq 1$,  $s_{av}$ shows a power-law increase $s_{av}\sim n^{\alpha'}$ and  from the fit to numerical data we obtain $\alpha'$(p=1)=0.26(1), $\alpha'$(p=3)=0.87(1), and $\alpha'$(p=5)=0.91(1).
\label{avsizelog}}
\vspace{-0.0cm}
\end{figure}

A similarly slow increase of  the giant component and of the giant community in the $p<1$ phase suggests that in this case communities are nearly identical to clusters. Our calculation of the average size of the community $s_{av}$ confirms such suggestion. As shown in Fig.~\ref{avsize}, the average community size is in a very good agreement with the analytical expression for the average cluster size $1/(1-p)$ \cite{newmanavsize}.
For $p\geq 1$, the average community size most likely has a power law increase $s\sim n^{\alpha'}$ (Fig.~\ref{avsizelog}). Similarly to $\alpha$, the exponent $\alpha'$ depends on $p$ and in particular, we estimate $\alpha'(p=1)\approx 0.26(1)$, $\alpha'(p=3)\approx 0.87(1)$, and $\alpha'(p=5)\approx 0.91(1)$.
Let us notice that $\alpha'(p=1)$ is nearly half of $\alpha(p=1)$. Moreover, for $p>1$ the numerical data show that $\alpha'(p)\approx \alpha(p)$.

\begin{figure}
\includegraphics[width=\columnwidth]{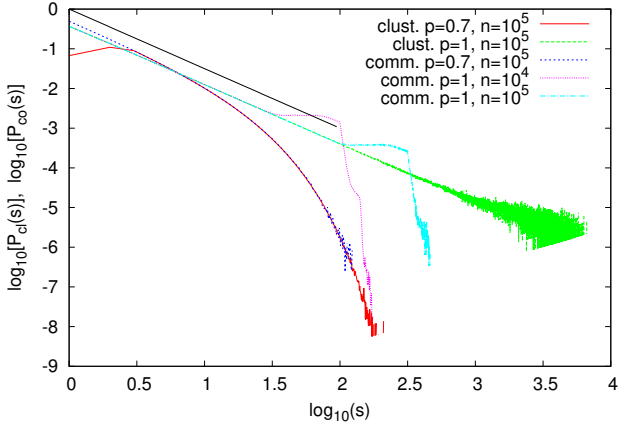}  \vspace{-0.8cm} 
\caption{(Color online) The distributions of size of clusters ($P_{cl}(s)$) and communities ($P_{co}(s)$), calculated for $p=1$ and $p=0.7$ and for $n=10^5$ (log-log scale). For $p=1$ both $P_{cl}(s)$ and $P_{co}(s)$ seem to have the expected power law decay $s^{-3/2}$ (thick black straight line) \cite{newmanavsize}. 
\label{distrib107}}
\vspace{-0.0cm}
\end{figure}

To have some further understanding of the community structure in random graphs, we measured the probability distribution of sizes $s$ of communities $P_{co}(s)$ and of clusters $P_{cl}(s)$. For $p<1$, clusters and communities are relatively small and, as expected, both distributions seem to have faster than a power-law decay (Fig.~\ref{distrib107}). Morever, the overlap of $P_{co}(s)$ and $P_{cl}(s)$ is yet another indication that communities in the non-percolating phase are nearly identical to clusters.
For $p=1$, both distributions seem to follow a power-law decay $s^{-3/2}$, which for the cluster size distribution $P_{cl}(s)$ is already well known \cite{newmanavsize}. Our results show that even at the critical point $p=1$, the distribution of communities $P_{co}(s)$ for any $s$ (and sufficiently large $n$) is almost identical to $P_{cl}(s)$ (Fig.~\ref{distrib107}). However, for a given $n$, the cutoffs of these distributions, as set by the size of the giant cluster and of the giant community, are different: the giant cluster scales as $n^{2/3}$ while the giant community is smaller and is likely to scale as $n^{1/2}$ (Fig.~\ref{log-giant}). From the asymptotic decay of $P_{co}(s)$ at the $p=1$, we can estimate the $n$-dependence of the average community size  $s_{av}\sim \int_0^{n^{1/2}} s P_{co}(s)ds\sim \int_0^{n^{1/2}} s^{-1/2}ds\sim n^{1/4}$, which agrees with our estimation $\alpha'=0.26(1)$. Generalizing, the giant community scaling as $n^{\alpha}$ will imply $s_{av}\sim n^{\alpha/2}$, i.e., $\alpha'=\alpha/2$.

\begin{figure}
\includegraphics[width=\columnwidth]{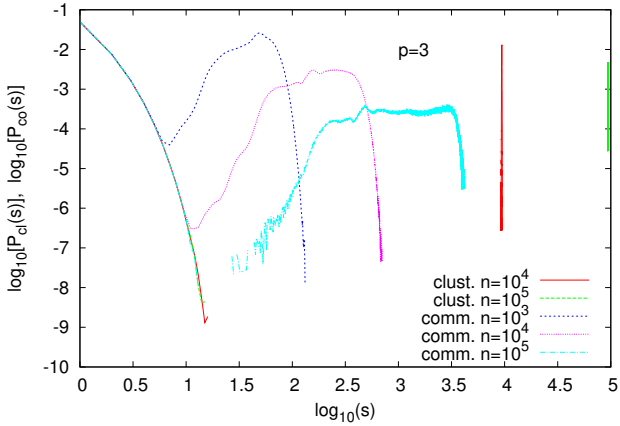}  \vspace{-0.8cm} 
\caption{(Color online) The distributions of the size of clusters ($P_{cl}(s)$) and communities ($P_{co}(s)$) calculated for $p=3$. The disconnected vertical parts of the $P_{cl}(s)$ on the right side correspond to spanning clusters.   
\label{distrib3}} 
\vspace{-0.0cm}
\end{figure}

\begin{figure}
\includegraphics[width=\columnwidth]{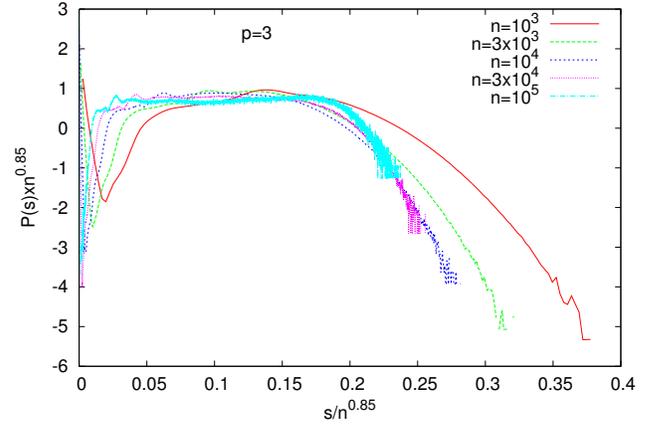}  \vspace{-0.8cm} 
\caption{(Color online) The distribution $P_{co}(s)$ rescaled with the characteristic giant community size $n^{0.85}$ calculated for $p=3$ and several values of $n$.  
\label{collapse}}
\vspace{-0.0cm}
\end{figure}

Much different distributions appear in the percolating phase ($p>1$). The existence of the percolating cluster divides $P_{cl}(s)$ into two basically separated parts (Fig.~\ref{distrib3}). The \mbox{large-$s$} part of the distribution corresponds to the spanning cluster and is separated from the  \mbox{small-$s$} part with a distance increasing as the size of the giant cluster ($\sim n$). For small $s$, the distribution $P_{co}(s)$ overlaps with $P_{cl}(s)$. For larger $s$, the distribution $P_{co}(s)$ seems to develop much wider maximum, which is not that much separated from the small-$s$ part. 
Such behaviour is related to the slower increase (with $n$) of the giant community, which most likely sets the scale of the characteristic size in the system. Rescaling $P_{co}(s)$ with $n^{0.85}$  (for $p=3$), we notice that the data for different $n$ approximately collapse (Fig.~\ref{collapse}). Moreover, the distribution becomes flat over a range $0<s/n^{0.85}\lesssim 0.2$ and the size at the upper limit of that interval $s\approx 0.2n^{0.85}$ is close to the size of the giant community. 
Let us notice that the size corresponding to the middle part of that interval, namely $0.1n^{0.85}$, is in a reasonably good agreement with the scaling of $s_{av}$. Such flat shape of the distribution strongly suggests that the scaling of $s_{g}$ and $s_{av}$ is governed by the same exponent, implying thus $\alpha=\alpha'$, in agreement with our numerical simulations (Fig.~\ref{log-giant}, Fig.~\ref{avsizelog}). Let us emphasize that the divergence of $s_{av}\sim n^{\alpha'}$ and $s_{g}\sim n^{\alpha}$ slower than the size of the graph $n$ together with the broad distribution of sizes $P_{co}(s)$ indicate that the regime $p>1$ with respect to community structure is critical.

\begin{figure}
\includegraphics[width=\columnwidth]{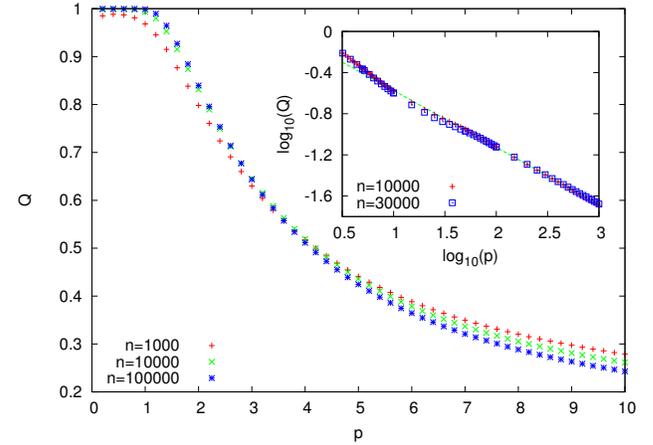}  \vspace{-0.7cm} 
\caption{(Color online)  The modularity $Q$ as a function of $p$. The inset presents the data for larger range of $p$ and suggests that for large $p$ the modularity decays as $p^{-0.55}$.
\label{modularity}}
\vspace{-0.3cm}
\end{figure}

As our final result in this section, we present the $p$-dependence of the modularity $Q$ (Fig.~\ref{modularity}). One can notice that for large $p$ the modularity $Q$ seems to have a power-law decay $\sim p^{-0.55}$. Our result is intermediate between the estimations $p^{-0.5}$ \cite{reichardt} and $p^{-2/3}$ \cite{amaral}, which were obtained by finding the ground-state configuration of a certain spin-glass model .
\section{Non-Greedy optimization of modularity}
The Newman greedy optimization of modularity is a very fast method suitable for finding the community structure of even very large networks. Preserving the basic algorithmic structure of this method (and thus its advantages), we introduce in this section a non-greedy optimization method. In the greedy version, one selects for amalgamation such pairs of communities that provide the largest increase of the modularity.
It is possible, however, that choosing other pairs of communities  will eventually lead to the community structure with larger modularity (than that found by the greedy optimization).  The essence of our method is to examine some other sequences of amalgamations,  where those resulting in the largest increase of modularity are preferred but others are also possible. In particular, we select a pair of communities to be amalgamated using a simple prescription: the larger the corresponding increase of modularity the better the chance of selecting a given pair. To be more specific, we use the roulette-wheel selection with the weight of an amalgamation, which will increase the modularity by $\Delta q$, given as $w(\Delta q)=1/(\Delta q_m-\Delta q+\varepsilon)$, where $\Delta q_{m}$ is the largest increase of modularity (available at a given stage of an algorithm) and $\varepsilon$ is a numerically determined parameter of the method \cite{comment-weight}. In the limit $\varepsilon\rightarrow 0$, the weight $w(\Delta q_m)$ dominates and our method becomes equivalent to the greedy optimization.  For positive $\varepsilon$, all weights are finite and other amalgamations (less greedy) might be selected. However, when $\varepsilon$ is too large, the algorithm is very noisy and does not drive the process to large-$Q$ solutions. Since the number of possible amalgamations is typically large, we used the $O(1)$ implementation of the roulette-wheel selection \cite{liplip2012}. Running the non-greedy algorithm several times, one might expect to find solutions better than that returned by the greedy optimization.

\begin{figure}
\includegraphics[width=\columnwidth]{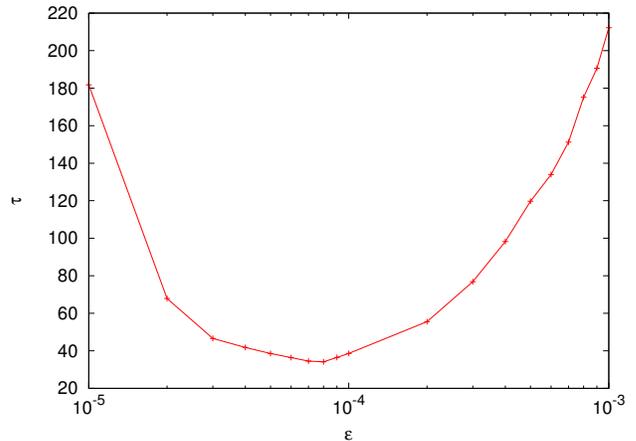}  \vspace{-0.7cm} 
\caption{(Color online)  Zachary's karate club network. The average number of runs $\tau$, which are needed to find the community structure with the largest modularity $Q=0.41979$, as a function of~$\varepsilon$.
\label{zachary}}
\vspace{-0.3cm}
\end{figure}

We tested our method with some simple benchmark networks. A well-known example is Zachary's karate club network \cite{zachary}, which has 34 nodes and for which the greedy optimization algorithm returns $Q=0.38067$ \cite{newman2004}. This is, however, only a suboptimal solution since the largest value, as obtained with simulated annealing  \cite{medus} or linear programming technique \cite{agarval}, equals $Q=0.41979$. Our algorithm also finds the community structure with $Q=0.41979$. To examine its performance we run the non-greedy algorithm for several values of $\varepsilon$, until the optimal solution (with $Q=0.41979$) was found. Repeating and averaging over $100$ trials, we calculated the average number of runs $\tau$, which are needed to find the optimal solution. Numerical calculations show (Fig.~\ref{zachary}) that in the vicinity of $\varepsilon=10^{-4}$ our algorithm easily finds the optimal solution. For larger $\varepsilon$, the algorithm amalgamates communities with less regard to the modularity increase while for smaller $\varepsilon$ it becomes too greedy. In both cases, finding the largest-modularity solution is more difficult.

\begin{figure}
\includegraphics[width=\columnwidth]{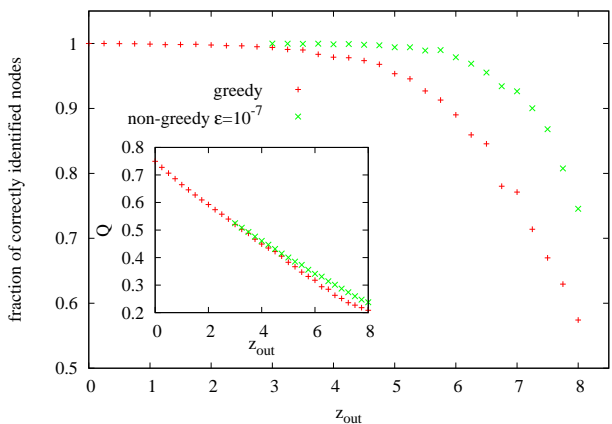}  \vspace{-0.7cm} 
\caption{(Color online)  The fraction of correctly identified nodes as a function of $z_{out}$ for the network having 1024 links and 128 nodes forming 4 groups, obtained using the greedy and non-greedy optimization. Each result is an average over 100 graphs and the non-greedy optimization was executed $10^3$ times for each graph. Inset shows the behaviour of the modularity $Q$ as a function of $z_{out}$.  
\label{4module}}
\vspace{-0.3cm}
\end{figure}

\begin{figure}
\includegraphics[width=\columnwidth]{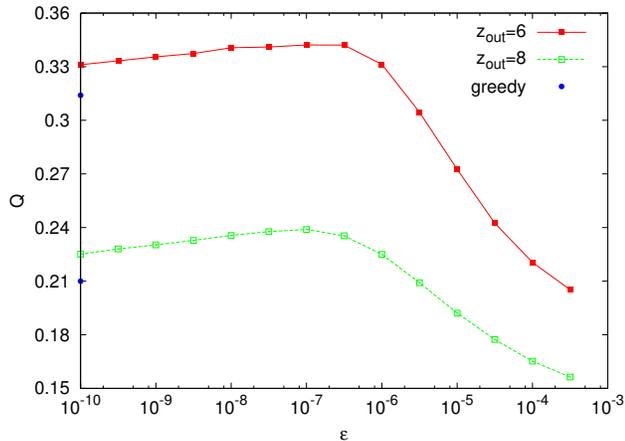}  \vspace{-0.7cm} 
\caption{(Color online)  The modularity $Q$ as a function of $\varepsilon$ for the 128-node network with 1024 links.  Circles indicate the values obtained by the greedy version ($\varepsilon=0$).
\label{modul68}}
\vspace{-0.3cm}
\end{figure}

We also tested our method on the network of 1024 links and 128 nodes that form 4 groups (of 32 nodes each). Each node has on average $z_{in}$  links with members of the same group and $z_{out}$ links with members of other groups ($z_{in}+z_{out}=16$). Also in this case our method outperforms the greedy optimization (Fig.~\ref{4module}). In particular, the non-greedy version returns a much larger fraction of correctly indentified nodes \cite{fraction} and a significantly larger modularity, especially for large $z_{out}$. Similar results for this set of networks were also found using the method of Extremal Optimization \cite{duch}.
In this case,  the method also has an optimal performance for a certain intermediate value of $\varepsilon$ (Fig.~\ref{modul68}). Let us notice that for $z_{out}=6$ and 8 the  optimal values of modularity (obtained for $\varepsilon\sim 10^{-7}$) are larger by more than 10\% than those found with the greedy algorithm. 

\begin{figure}
\includegraphics[width=\columnwidth]{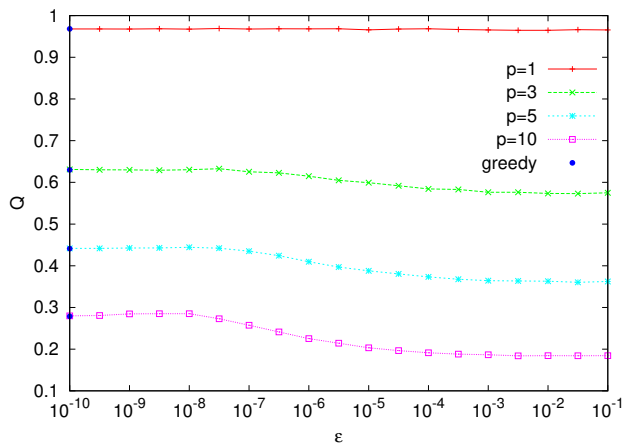}  \vspace{-0.7cm} 
\caption{(Color online)  The modularity $Q$ as a function of $\varepsilon$ for $n=1000$ random graph. Each result is an average over 100 graphs and non-greedy optimization was executed $10^3$ times for each graph.
Circles indicate the values obtained by the greedy version ($\varepsilon=0$).
\label{rand1000modul}}
\vspace{-0.3cm}
\end{figure}

Having tested the method for some well-known examples, we applied the non-greedy optimization to random graphs. Our calculations show, however, that in this case the non-greedy method offers little or maybe no improvement. Indeed, the largest values of modularity seem to be the same as those previously obtained by the greedy method (Fig.~\ref{rand1000modul}). We present only $n=1000$ data but calculations for other values of $n$ show a similar behaviour.

\begin{figure}
\includegraphics[width=\columnwidth]{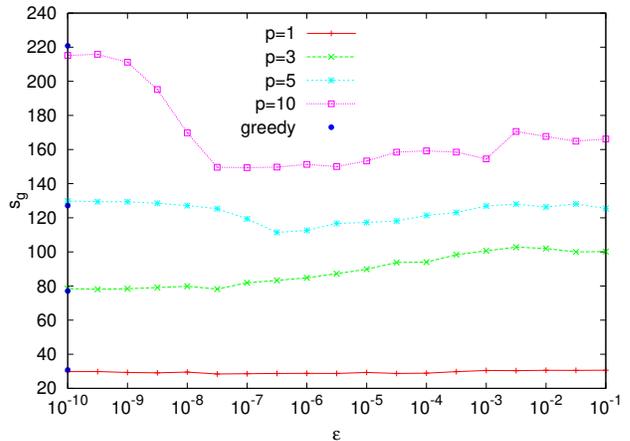}  \vspace{-0.7cm} 
\caption{(Color online)  The average size of the giant community as a function of $\varepsilon$ for $n=1000$ random graphs. Each result is an average over 100 graphs and the non-greedy optimization was executed $10^3$ times for each graph.
Circles indicate the values obtained by the greedy version ($\varepsilon=0$).
\label{rand1000max}}
\end{figure}

Upon changing $\varepsilon$, the size of the giant community $s_g$ exhibits some variability (Fig.~\ref{rand1000max}). However, in the regime of small $\varepsilon$, i.e., with the largest modularity (Fig.~\ref{rand1000modul}), the values of $s_g$ obtained by the non-greedy and greedy methods nearly overlap. Thus, we expect that the results obtained using the greedy method, which were reported in section III, are nearly optimal since the non-greedy method does not find solutions of larger modularity.

\section{Conclusions}
In summary, we examined the community structure in random graphs and compared it with their cluster properties. While below the percolation threshold, clusters and communities are nearly the same, at the percolation threshold and especially above it, the communities are much smaller than the clusters. There are some interesting consequences, especially above the percolation threshold, where, as we show, contrary to cluster properties there is no spanning community. Since the average size of communities diverges but slower then the size of a graph and there is a broad distribution of sizes of communities, the regime above the percolation threshold, with respect to the community structure, might be considered critical. An open question is whether the above described critical behaviour, which we  detected numerically for moderately dense graphs, will persist in denser networks. Alternatively, one might expect yet another threshold, above which a spanning community will form.

To determine the community structure, we used the Newman method based on the greedy optimization of the modularity. 
We checked, however, that nearly the same results are obtained by the non-greedy optimization method.

Introducing the non-greedy optimization method presents an additional objective of the present paper. The method preserves an efficient algorithmic structure of the greedy optimization  but relaxes the requirement that always those pairs of communities are amalgamated, which yield the largest possible increase of modularity. Instead, some other amalgamations take place and  this eventually might lead to determination of a community structure with larger modularity, as we demonstrated on some examples. We expect that the method might be further improved.  One possibility is to continuously change the parameter $\varepsilon$ so that, for example, at the beginning (for large $\varepsilon$) various amalgamations would be possible---also those leading to a small increase (or even decrease) of modularity. However, during the run, such steps would be gradually suppressed and at the end of the process ($\varepsilon\rightarrow 0$) only the largest increase would be admissible. Such a procedure would clearly resemble the simulated annealing technique but an examination of its efficiency is left for the future.   

Acknowledgments: D.L. is supported by NCN grant 2011/01/B/HS2/01293 and A.L. is supported by NCN grant 2013/09/B/ST6/02277


\begin{thebibliography}{}
\bibitem{comment-weight} In principle some amalgamations might  decrease the modularity but we assume that they are forbidden.
\bibitem{fraction} To calculate the fraction of correctly identified nodes, we examined all 
injective assignments of communites determined by our algorithms (either greedy or non-greedy) to the four groups of nodes and selected that with the largest overlap. For each graph, we executed the non-greedy algorithm $10^3$ times and analyzed only the community structure with the largest modularity. Each data point in Fig.~\ref{4module} is an average over 100 graphs.
\end{thebibliography}
\end {document}